\documentclass[aps,prl,twocolumn]{revtex4-2}
\usepackage{amsmath,amsthm,amssymb}

\newcommand{\proj}{P}
\newcommand{\projerr}{P^\perp}
\DeclareMathOperator{\dd}{d}
\DeclareMathOperator{\var}{var}
\DeclareMathOperator\atanh{atanh}

\newcommand{\tdv}[1]{\tfrac{\dd}{\dd\!#1}}
\newcommand{\ev}[1]{\langle#1\rangle}

\newcommand{\dyad}[2]{|#1\rangle\kern-0.6ex\langle#2|}
\DeclareMathOperator{\tr}{tr}
\newcommand{\lindblad}{\mathcal{L}}
\newcommand{\bigO}[1]{\mathcal{O}(#1)}
\newcommand{\textsum}[1]{{\textstyle\sum_{#1}}}

\begin{document}

\title{A Universal Constraint on Computational Rates in Physical Systems}
\author{Hannah Earley}
\email{h.earley@damtp.cam.ac.uk}
\affiliation{Department of Applied Mathematics and Theoretical Physics\\University of Cambridge}
\begin{abstract}
  Conventional computing has many sources of heat dissipation, but one of these---the Landauer limit---poses a fundamental lower bound of 1 bit of entropy per bit erased.
  `Reversible Computing' avoids this source of dissipation, but is dissipationless computation possible?
  In this paper, a general proof is given for open quantum systems showing that a computer thermally coupled to its environment will necessarily dissipate entropy (and hence heat).
  Specifically, a lower bound is obtained that corresponds to the adiabatic regime, in which the amount of entropy dissipated per computational operation is proportional to the rate of computation.
\end{abstract}
\maketitle

\section{Introduction}

In 1961, Rolf Landauer~\cite{landauer-limit} showed that information `erasure' has a fundamental entropic cost, of $\Delta S\ge \Delta I$ where $\Delta I$ is the amount of Shannon information being erased, and in units in which Boltzmann's constant $k_B=1$.
This imposes a minimum heat dissipation on any conventional computer, due to the reliance of conventional models of computation on irreversibility and information erasure.
Soon, Charles Bennett~\cite{bennett-tm} and others demonstrated that \emph{reversible} computation---computation which is logically reversible, and hence is not permitted to erase information---was possible.
Moreover, Bennett showed that such computation was just as powerful as irreversible computation.
A notable early contribution was that of Fredkin and Toffoli~\cite{fredkin-conlog} with their model of Ballistic Billiard-Ball Computation: a reversible computer constructed from ideal elastic billiard balls and ideal rigid walls, and subject to classical mechanics.

In principle, ballistic reversible computation would dissipate no energy.
For Billiard-Ball Computation, energy need only be supplied at the beginning in the form of the balls' kinetic energies, and may be recovered at the end.
Unfortunately, Bennett~\cite{bennett-rev} showed that this model was highly susceptible to fluctuations: even very weak gravitational influences, such as from turbulence in the atmospheres of nearby stars, would be sufficient to thermalize the motion of the system within a hundred-or-so collisions.
Further work to study the costs of reversible computing has been completed by Frank~\cite{frank-thesis}, Levitin and Toffoli~\cite{levitin-rc-cost}, and Pidaparthi and Lent~\cite{pidaparthi-lent}.

In this paper, we analyze the amortized entropy cost of reversible computers at a given finite temperature, and the resulting upper bound this places on the rate of computation for a system with a given maximum rate of heat dissipation and supply of free energy.
We thus show a universal upper bound on the rate of computation in a given region of space, corresponding to the adiabatic case analyzed in Frank's thesis~\cite{frank-thesis}.
The paper starts by setting up a generic quantum model of a reversible computer.
We then evaluate it in two regimes:
  (1) the first regime applies when the thermal influence of the external environment is negligible, in which case the Quantum Zeno Effect can be exploited;
  (2) the second regime applies when the external environment is significant.
Both regimes turn out to yield the same asymptotic scaling result, though with a qualitative difference in the permitted internal structure of the computational system.

\section{Setup}

Throughout this paper, we will use a simple open quantum system approach.
A computational system $\mathcal S$ consists of a computational part under our control, $\mathcal C$, and an external environment outside of our control, $\mathcal E$.
Specifically, we may perform arbitrary operations on $\mathcal C$ at arbitrary times, but not on $\mathcal E$.
Furthermore, we assume there are no exploitable correlations between $\mathcal C$ and $\mathcal E$;
  that is, after a brief time following an interaction between $\mathcal C$ and $\mathcal E$, the environment rapidly relaxes such that we can factorize the system state as $\rho_{\mathcal S}=\rho_{\mathcal C} \otimes \rho_{\mathcal E}$.
Throughout its operation, $\mathcal C$ will produce entropy, which will concomitantly be exported to the environment $\mathcal E$.
Following standard notation~\cite{entropy-prod}, the rate of change of entropy $\dot S$ is given by the difference between the rate of entropy production $\dot\Sigma$ and the entropy flow rate $\dot\Phi$.
Amortized over time, these are balanced such that the entropy of $\mathcal C$ remains constant: $\dot S=0$.

In the absence of $\mathcal E$, the computational part $\mathcal C$ evolves unitarily under a Hamiltonian $H=H_0 + U(t)$.
$H_0$ corresponds to the ballistic computational evolution of the computational state.
$U(t)$ is a time-variable, zero-mean, Gaussian potential arising from the finite temperature of $\mathcal C$; it is adiabatic, in that it operates solely on $\rho_{\mathcal C}$.
Interactions between $\mathcal C$ and $\mathcal E$ are governed by an interaction Hamiltonian $H_I$, which we will model with a standard Gorini–Kossakowski–Sudarshan–Lindblad (GKSL) approach.
That is, the computational state $\rho=\rho_{\mathcal C}$ evolves non-unitarily via the completely-positive trace-preserving GKSL equation
\begin{align*}
  \dot\rho &= i[\rho, H] + \lindblad\rho,
\end{align*}
in units in which $\hbar=1$, and where $\lindblad\rho = \sum_i\gamma_i(L_i\rho L_i^\dagger-\frac12\{L_i^\dagger L_i, \rho\})$ is the Lindblad operator with strengths $\gamma_i$ and jump operators $L_i$.

For sustained operation of the computer, errors induced by the thermal coupling of $U$ and $\lindblad$ must be corrected.
For long runtimes, these must be corrected regularly to prevent thermalization of the computational state and invalid operation.
To proceed, consider the probability of an error occurring, $\delta p$, in a short time, $\delta t$.
Let the state space of $\mathcal C$ be partitioned by two projectors, $\proj$ and $\projerr$ corresponding to a currently valid and currently invalid state respectively.
The system need not necessarily reset to some ideal unique state in each correction, and $\proj$ takes into account the range of permissible states to which we may reset.
We will use the fact that $\proj\rho=\rho\proj=\rho$ and $\projerr\rho=\rho\projerr=0$ when $\rho$ is in a valid state.
The probability of error is given by measuring $\rho(t+\delta t)$ in this basis, i.e.\ $\delta p=\tr[\projerr\rho(t+\delta t)]$.
Letting $\rho=\rho(t)$ and expanding in $\delta t$ up to second order,
\begin{align*}
  \delta p &= \tr[\projerr(\rho + \delta t\dot\rho + \tfrac12\delta t^2\ddot\rho)] + \bigO{\delta t^3}.
\end{align*}
We will need
\begin{align*}
  \ddot\rho &= i([\dot\rho, H] + [\rho,\dot H]) + \tdv{t}\lindblad\rho \\
    &= -\rho H^2 + 2H\rho H - H^2\rho + i[\lindblad\rho,H] + i[\rho,\dot H] + \tdv{t}\lindblad\rho
\end{align*}
from which we find
\begin{align*}
  \delta p &= \tr[\projerr(\delta t \lindblad\rho + \delta t^2 (H\rho H + \tfrac12 i[\lindblad\rho,H] + \tfrac12\tdv{t}\lindblad\rho))]
\end{align*}
up to $\delta t^2$.
Note that, whilst we are framing this in terms of classical reversible computation and error correction, the above analysis and following results should also hold for quantum computation and quantum error correction schemes.

There are now two different regimes depending on the overall strength of $\lindblad$.

\section{Quantum Zeno Evolution}

The first order term in $\delta p$ is $\delta t\tr[\rho \sum_i\gamma_i L_i^\dagger\projerr L_i]$.
If the Lindbladian is weak, such that combined jump operator sum scales as $\bigO{\delta t}$, then $\delta p=\bigO{\delta t^2}$.
This scaling of $\delta p$ is characteristic of the Quantum Zeno Effect, in which quantum evolution can be strongly suppressed by frequent measurements.
This is because the rate of errors is $\dot p=\bigO{\delta t}$, which goes to 0 as the measurement interval goes to 0.
In practice, there is a lower bound on the \emph{effective} measurement interval given by the Margolus-Levitin limit~\cite{margolus-levitin} on the rate of quantum evolution.

In this case, we obtain (using $\rho=\rho\proj$)
\begin{align*}
  \delta p &= \delta t^2 \tr[\rho\proj( H\projerr H + i\textsum{i}(\gamma_i/\delta t) L_i^\dagger\projerr L_i )].
\end{align*}
Now, writing the operators $H_0$, $U$, $L_i$ in block-matrix form as $(\begin{smallmatrix}H_{00} & H_{01} \\ H_{10} & H_{11}\end{smallmatrix})$ etc.\ in the basis $(\proj,\projerr)$, $\delta p$ reduces to
\begin{align*}
  \delta p &= \delta t^2 \tr[\rho_{00}(U_{01}U_{01}^\dagger + i\textsum{i}(\gamma_i/\delta t)L_{i,01}^\dagger L_{i,01} )];
\end{align*}
i.e.\ the probability of error is induced solely by the off-diagonal connections between the valid and invalid states in the thermal operators; it is independent of $H_0$.
We aggregate these thermal operators as an effective thermal potential $U'$, $\delta p = \delta t^2 \tr[\rho U'\projerr U'^\dagger] = \delta t^2 \ev{U'\projerr U'^\dagger}$.

When an error occurs, the state of the system becomes less certain and there is a corresponding increase in entropy.
For a computational system $\mathcal C$, we can assume that there will be a set of $M$ possible valid computational states with associated projectors $\mathcal P=\{P_i\}$, such that our $\proj\in\mathcal P$.
The remaining states of the system we deem non-computational, and to these we associate the projector $P_\varnothing$.
Then, $\projerr=\openone - \proj$ and $\openone = P_\varnothing + \sum_i P_i$.
Now, it is reasonable to assume that the energy spectra of each of the $P_i$ are very similar, as otherwise the computer would prefer certain computations to others (and the higher energy ones would have a higher likelihood of error); however, we impose no such constraint on the $P_\varnothing$.
Without loss of generality, we assume the $P_\varnothing$ states are energetically inaccessible as this weakens the bound and so does not invalidate the final result.
With the assumption of the $P_i$ being energetically similar, and that $U'$ is uncorrelated with and ignorant of the system dynamics, we can say that $U'$ will be equally likely to send $\rho$ to any of the $P_i$.
The increase in entropy production $\Sigma$ will then be given by
\begin{align*}
  \delta\Sigma &= [-\delta p\log\delta p - (1-\delta p)\log(1-\delta p) +{} \\
    &\quad\quad\quad \delta p (S_0 + \log(M-1)) + (1-\delta p) S_0] - [S_0] \\
    &= \delta p (\log(M-1) + 1 - \log\delta p) + \bigO{\delta p^2},
\end{align*}
where $S_0$ is the intra-state entropy (i.e.\ the entropy of the distribution of states within $\proj$).
If $S_0$ is non-maximal, then it will tend to increase between measurements, but without loss of generality one can simply expand the set of projectors $\mathcal P$ (by subdividing the $P_i$) until $S_0$ is maximal;
  otherwise the increase in intra-state entropy between measurements would not be subject to the Quantum Zeno Effect, dominating the entropy production and effectively reducing the computation rate.

To be useful for computation, the system will need to have many possible computational states, i.e.\ $M\gg1$.
$M$ is maximized when the computational state decomposes into a set of $m$ independent digital subsystems, each taking on one of at least $g_i=2$ possible values; that is, $M=g^m$ where $g$ is the geometric mean of the $g_i$.
We can then write $\delta\Sigma\ge\delta p \cdot m \log g$, or $\dot\Sigma\ge \delta t\ev{U'\projerr U'^\dagger}m\log g$.

To evaluate the contribution of the effective potential $U'$, we make use of the decomposition into $m$ independent digital subsystems.
These subsystems may themselves be composed of $n_i$ \emph{primitive} quantum subsystems/particles.
Each particle $ij$ has an effective temperature $T_{ij}$ under $U'$, and $U'$ acts independently on each due to the assumption of the potential being ignorant of, and uncorrelated with, the computational system $\mathcal C$.
As the time-average of the potential is vanishing, $\overline{\ev{U'\projerr U'^\dagger}}=(\tr\projerr/\tr\openone)\overline{\var U'}\approx \sum_{ij}T_{ij}^2=mnT^2$, where we have used the independence of $U'$ from the details of the computational system to take out $\projerr$ as a factor $\tr\projerr/\tr\openone$, and $M\gg1$ to take $\tr\projerr\approx\tr\openone$.
$n$ and $T$ are the averages of the $n_i$ and $T_{ij}$ respectively.

This gives $\dot\Sigma\ge\delta t\cdot nm^2 T^2\log g$.
Now, $\delta t=1/\nu_Z$ where $\nu_Z$ is the rate of (Zeno) measurements.
By the Margolus-Levitin~\cite{margolus-levitin} limit on quantum transition rates, the computation rate is bounded by $\nu_C\le nm\varepsilon/n\pi$ where $\varepsilon$ is the average energy per particle.
Note that we divide by $n$ because the $n_i$ particles in each computational subsystem are redundant and so do not meaningfully contribute to the rate of computation.
Hence,
\begin{align*}
  \dot\Sigma &\ge \frac{\nu_C^2}{\nu_Z} \underbrace{\frac{n\pi^2T^2}{\varepsilon^2} \log g}_\eta.
\end{align*}
Simple rearrangement gives an upper bound to the rate of computation, $\nu_C\le\sqrt{\dot\Sigma\nu_Z}/\eta$.
By the Margolus-Levitin limit, we have $\nu_Z\le E/\pi$ and hence the scaling law can be written in terms of the rate of heat dissipation and the mass-energy of the system, $\nu_C\lesssim\sqrt{\dot\Sigma E}$.
To see that this is characteristic of the adiabatic regime of reversible computing, we rearrange to find the dissipation-delay product.
The dissipation per operation is given by $\Delta\Sigma=\dot\Sigma/\nu_C$, and the delay per operation is simply $\tau_C=1/\nu_C$, and so we have $\Delta\Sigma\cdot\tau_C \ge\pi\eta/E$, a constant.
That is, the per-operation dissipation scales inversely with delay.

\subsection{Composite Systems}

The quadratic dependence of $\dot\Sigma$ on $\nu_C$ makes it unclear that the same bound holds for a composite system in which $\mathcal C$ is divided into $N$ independently corrected subsystems.
To show that it does, we maximize the total computational rate---with respect to the individual rates $\nu_{C,i}$---given the total entropy production rate bound by introducing the Lagrangian multiplier $\lambda$:
\begin{align*}
  \Lambda &= \sum_i \nu_{C,i} - \frac{1}{2\lambda} \sum_i \frac{\nu_{C,i}^2}{\nu_{Z,i}} \eta_i, \\
  0 &= 1 - \frac{1}{\lambda} \frac{\nu_{C,i}}{\nu_{Z,i}} \eta_i,
\end{align*}
i.e.\ $\dot\Sigma_i/\nu_{C,i} \ge \lambda$ is constant.
With a bit of rearrangement, we find
\begin{align*}
  \lambda \nu_{Z,i} &= \nu_{C,i} \eta_i, \\
  \lambda \nu_Z &= \nu_C \ev{\eta_i}_{\nu_{C,i}},
\end{align*}
where in the last line we have summed over $i$ and the expectation value is weighted by the fractional computational rates of each subcomponent.
This recovers the same bound between entropy production rate and computational rate,
\begin{align*}
  \dot\Sigma_i \ge \frac{\nu_{C,i}^2}{\nu_{Z,i}} \eta_i = \lambda\nu_{C,i}
  \implies
  \dot\Sigma \ge \lambda\nu_C = \frac{\nu_C^2}{\nu_Z} \bar\eta.
\end{align*}

\section{Strong Lindbladian Regime}

If the Lindbladian is strong, such that $\delta p=\bigO{\delta t}$, then we instead obtain $\dot\Sigma \ge \ev{U'\projerr U'^\dagger} m \log g=nm^2T^2\log g$, i.e. without the $\delta t=1/\nu_Z$ contribution (and with $U$ not contributing to $U'$).
Here the $T$ corresponds to the effective average `temperature' of the Lindbladian interaction.
This gives $\dot\Sigma_i\ge\nu_{C,i}^2\eta_i$.
Applying the same variational approach as above, we obtain $\dot\Sigma \ge (\nu_C^2/N)\bar\eta$.
That is, instead of $\nu_Z$ we get the number of subsystems of $\mathcal C$, $N$.
This yields a bound $\nu_C \lesssim \sqrt{\dot\Sigma N}$

Unlike in the Zeno case, to maximize this rate of computation, $\mathcal C$ must be maximally subdivided.
Assuming a minimal useful computational subsystem has energy $E_1$, $N$ is maximized by $E/E_1$ where $E$ is the total energy of the system.
This recovers the same form of bound as the Zeno case.

\section{Classical Realization}

This bound can also be saturated by a classical system.
As a minimal example, we consider a simple Chemical Reaction Network (CRN) capable of performing computation.
A CRN consists of a set of species, $\{X_i:i\}$, and a set of reactions between them, $\{\sum\nu_{ij}X_i\to\sum\nu_{ij}'X_i:j\}$, where $\nu_{ij}$ and $\nu_{ij}'$ are the \emph{stoichiometries} of the `reactants' and `products' respectively.
Consider the CRN consisting of the computational species $\{C_i:i\in\mathbb Z\}$, representing the sequential computational states encountered during some program, and the `bias' species $\{\oplus,\ominus\}$.
For simplicity, we assume there is no initial or final state, and so $i$ ranges over $\mathbb Z$.
The bias species drive the computation forward or backward via the pair of (reversible) reactions $\oplus+C_i \leftrightarrow \ominus+C_{i+1}$.

It can be readily seen that the total number of $C$ particles and the total number of bias particles each remains invariant.
The fraction of bias species that are $\oplus$ or $\ominus$ may be modelled as a Continuous-Time Markov Chain (CTMC).
The entropy production for a CTMC is given by~\cite{entropy-prod-ctmc}
\begin{align*}
  \dot\Sigma &= \frac12 \sum_{ij} (\nu_{i\mapsto j} - \nu_{j\mapsto i}) \log\frac{\nu_{i\mapsto j}}{\nu_{j\mapsto i}} \\
    &= \sum_{ij} \nu_{ij} \beta_{ij} \atanh \beta_{ij}
\end{align*}
where $\nu_{i\mapsto j}=p_i\Gamma_{i\mapsto j}$ is the rate of the reaction $i\mapsto j$, $p_i$ is the proportion of state $i$, $\Gamma_{i\mapsto j}$ is the transition rate from state $i$ to $j$, $\nu_{ij}=\nu_{i\mapsto j} + \nu_{j\mapsto i}$ is the gross reaction rate between states $i$ and $j$, and $\beta_{ij}=\frac{\nu_{i\mapsto j} - \nu_{j\mapsto i}}{\nu_{i\mapsto j} + \nu_{j\mapsto i}}\in[-1,1]$ is the `bias' of the reaction.

We assume microscopic reversibility and detailed balance such that at equilibrium $\nu_{\oplus\to\ominus}=\nu_{\ominus\to\oplus}$.
We also assume that no computation happens at equilibrium, giving $p^{\text{eq.}}_\oplus=p^{\text{eq.}}_\ominus=\frac12$ and hence $\Gamma_{\oplus\to\ominus}=\Gamma_{\ominus\to\oplus}=\Gamma$.
This gives $\dot\Sigma=\Gamma\beta\atanh\beta$ where $\beta=\beta_{\oplus,\ominus}=p_\oplus-p_\ominus$.

For $\beta\sim\bigO{1}$, $\atanh\beta$ grows logarithmically in $1-\beta$ and diverges to infinity: we lose adiabaticity.
Contrastingly, in the limit $\beta\ll 1$ we have $\dot\Sigma=\Gamma\beta^2 + \bigO{\beta^4}$.
Considering the entire system, we have $\dot\Sigma=N\Gamma\beta^2$ where $N$ is the number of bias particles and where $\Gamma$ takes into account the second order reaction rate with the $C_i$.
The rate of computation is given by $\nu_C=\Gamma\beta N$.
This gives $\nu_C^2=\Gamma N\dot\Sigma$, recovering the adiabatic scaling law.

\section{Discussion}

The two regimes correspond to each of the two contributions to the value of $\eta$, that from $U$ and that from $\lindblad$, dominating.
Whilst each regime gives the same scaling law (as indeed do intermediate regimes), there is a qualitative difference between the two.

In the Zeno regime, the system may be subdivided as little or as much as desired.
If $M=1$ then the system is a single serial computer, operating at $\nu_{C,1}\lesssim\sqrt{\dot\Sigma E}$.
If $M=E/E_1$, each subsystem operates at $\nu_{C,1}\lesssim\sqrt{\dot\Sigma\smash{{}/{}}E}$ whilst the entire computer $\mathcal C$ operates at $\nu_C\lesssim\sqrt{\dot\Sigma E}$.
It can hence be seen that as subdivision increases, the individual computational subsystems operate at slower speeds whilst the net computation rate remains the same.
Contrastingly the strong-Lindbladian regime has no choice in the amount of subdivision: computation is maximized by maximally subdividing the system.

We can further understand these operating speeds by obtaining geometric bounds on $\dot\Sigma$ and $E$.
We assume a fixed method to construct a computer, such that a computer $\mathcal C_2$ which is double the size of $\mathcal C$ is roughly equivalent to two copies of $\mathcal C$.
It follows that the mass-energy $E$ scales with the occupied computational volume $V$.
Meanwhile the computer must exchange entropy and free energy with its environment.
Assume that this is performed by some device which can transfer heat/free energy at some maximum rate per unit area.
The convex bounding surface of $\mathcal C$ is where this flux is most concentrated, as it has the minimum bounding surface area $A$.
Hence $\dot\Sigma\sim A$.
Thus the scaling law for the net rate of computation can be written $\nu_C\lesssim\sqrt{AV}$, and the rate of individual subsystems in a maximally subdivided computer may be written $\nu_{C,1}\lesssim\sqrt{A/V}$.
For a spherical region, $A\sim r^2$ and $V\sim r^3$ giving $\nu_C\lesssim r^{5/2}$ and $\nu_{C,1}\lesssim r^{-1/2}$.
Hence the individual rate of computation asymptotically vanishes as the size of $\mathcal C$ increases, whilst the net rate of computation grows.

In an idealized system, the influence of the environment $\mathcal E$ may be considered localized to the boundary and hence to scale with $A$.
One may then think that a sufficiently large computer can be run in the Zeno regime by suppressing the contribution of $\lindblad$ from $\eta$.
However this assumes that either
  (1) we may operate $\mathcal C$ as a single undivided entity, performing a projective measurement $P$ monolithically on the entire system; or
  (2) computational subsystems in the Zeno regime are thermally isolated from each other.
If the latter constraint is violated, then the combined boundaries of each subsystem contribute to the effective size of $\mathcal L$.
As both constraints are unlikely to be realized, it is probable that the strong-Lindbladian regime will always be relevant.
Hence it may be impractical to exploit the Quantum Zeno effect in maximally efficient reversible computers.

Compared to irreversible computing, this adiabatic scaling provides asymptotically superior computational performance.
The relevant law for irreversible computers is $\nu_C\lesssim \dot\Sigma/\Delta I$ where $\Delta I$ is the amount of information erased per computational operation.
Hence, if $E$ scales asymptotically faster than $\dot\Sigma$ then the rate of a reversible computer will asymptotically beat that for an irreversible computer.
Indeed, for the geometric limits above we obtain $\nu_C^{\text{rev.}}\lesssim r^{5/2}$ and $\nu_C^{\text{irrev.}}\lesssim r^2$, showing a scaling advantage of $r^{1/2}$.

Levitin and Toffoli~\cite{levitin-rc-cost} obtain similar results, in particular finding $\dot\Sigma\sim\nu_C^2$.
However, their full expression is $\dot\Sigma\sim \Delta p\,\nu_C^2/N$ where $\Delta p$ is the probability of error per computational step, and $N$ is the number of computational states with $\nu_C\propto N$.
Note that this value of $N$ differs from our use of $N$ to refer to the number of computational subsystems in a composite system.
As a result, their expression reduces to $\dot\Sigma\sim \Delta p\,\nu_C$, expressing a seemingly linear dependence of the rate of entropy generation on the rate of computation.
Where our methodology differs is in studying the value of $\Delta p$ to find how it depends on factors such as $\nu_C$.
In contrast, Levitin and Toffoli assume $\Delta p$ is independent of computational rate.

In related work, Pidaparthi and Lent~\cite{pidaparthi-lent} analyze dissipation for the concrete system of a Quantum Cellular Automaton.
They observe, in an isolated system, that dissipation can decay exponentially with the delay---rather than linearly as with normal adiabatic processes.
Their analysis helps to explain the origin of linear adiabaticity in open systems.
It also describes the behaviour at fast switching times, where the strict linear relationship is not observed.
As this regime is still subject to a lower bound it does not undermine our asymptotic scaling limits, but it does demonstrate interesting behaviour that is relevant to engineering real-world reversible systems at fast switching speeds.

\section{Acknowledgements}

The author would like to acknowledge the support of her PhD supervisor, Gos Micklem, and useful feedback from Mike Frank.
This work was supported by the Engineering and Physical Sciences Research Council, project reference 1781682.

\bibliography{references}


\end{document}